\documentclass[prc,aps,amsmath,amssymb,nofootinbib,superscriptaddress,12pt]{revtex4-2}
\usepackage{epsfig}
\usepackage[bookmarksnumbered,bookmarksopen,colorlinks,citecolor=blue,linkcolor=blue]{hyperref}
\usepackage{amsmath}
\usepackage{amssymb}
\usepackage{tipa}

\begin{document}

    \title{ Effective nucleus-nucleus potentials for heavy-ion fusion reactions }
 
    \author{Ning Wang}
    \email{wangning@gxnu.edu.cn}\affiliation{ Department of Physics,
        Guangxi Normal University, Guilin 541004, People's Republic of
        China }
    \affiliation{ Guangxi Key Laboratory of Nuclear Physics and Technology, Guilin 541004, People's Republic of
        China }

\author{Jinming Chen}
\affiliation{ Department of Physics, Guangxi Normal University, Guilin 541004, People's Republic of
	China }

\author{Min Liu}
\email{liumin@gxnu.edu.cn}\affiliation{ Department of Physics,
	Guangxi Normal University, Guilin 541004, People's Republic of
	China }
\affiliation{ Guangxi Key Laboratory of Nuclear Physics and Technology, Guilin 541004, People's Republic of
	China }

    \begin{abstract}
        Based on the Skyrme energy density functional and the reaction $Q$-value, we propose an effective nucleus-nucleus potential for describing the capture barrier in heavy-ion fusion processes. The 443 extracted barrier heights are well reproduced with a root-mean-square (rms) error of 1.53 MeV and the rms deviations with respect to 144 TDHF capture barrier heights is only 1.05 MeV. Together with the Siwek-Wilczy\'{n}ski formula in which the three parameters are determined by the proposed effective potentials, the measured capture cross sections at energies around the barriers can be reasonably well reproduced for a series of fusion reactions induced by not only nearly spherical nuclei but also the nuclei with large deformations such as $^{154}$Sm and $^{238}$U. The shallow capture pockets and small values of the average barrier radii play a role in the reduction of the capture cross sections for $^{52,54}$Cr and $^{64}$Ni induced reactions which are related to the synthesis of new super-heavy nuclei. 
    \end{abstract}

     \maketitle

\newpage
  
   \begin{center}
  	\textbf{ I. INTRODUCTION }\\
  \end{center}
  
  The investigation of heavy-ion fusion reactions holds significant importance, not only for the synthesis of new super-heavy nuclei (SHN) \cite{Hof00,Hof04,Mori04a,Ogan06,Ogan10,Ogan15,Ogan22,Sob18,Mori20,Pomo18,Adam04,Wang15,WangZ,Pei24} and extremely proton-rich nuclei \cite{Zhang19,Zhang21,Zhou21}, but also for the exploration of nuclear structures \cite{Stok78, Sarg11, Das98, Jia14, Yang23}. In the case of fusion reactions involving light and intermediate nuclei, approaches such as fusion coupled channel calculations \cite{Hag99,Dasso87,Das98} or empirical barrier distribution methods \cite{Zag01,SW04,liumin,Wang09,Wangbing,Jiang22} are adopted to calculate capture (fusion) cross sections. These calculations are often based on static or dynamic nuclear potentials \cite{Wong73,Bass74,Bass80,BW91,Gup92,Wen22,Wang08}. The static potentials are typically characterized by using models such as the liquid-drop model, energy density functional, or double-folding concept together with the sudden approximation. A precise characterization of the nucleus-nucleus potential, particularly at short distances, is essential for comprehending the fusion mechanism. 

In Ref. \cite{liumin}, a static entrance channel nucleus-nucleus potential is proposed to describe heavy-ion fusion reactions, using the Skyrme energy density functional  \cite{Vau72} combining the extended Thomas-Fermi (ETF) approach \cite{Brack,Bart02,Deni02} and the sudden approximation for densities. By introducing an empirical barrier distribution composed of a combination of two Gaussian (2G) functions to account for the dynamic effects in fusion processes, the most probable barrier heights $V_B\approx 0.946 B_0$ (with the frozen barrier height $B_0$) and fusion excitation functions for various reactions can be reasonably well described \cite{liumin,Wang09,Wang08}. Although certain measured fusion cross sections can be accurately reproduced, the realistic nucleus-nucleus potential in fusion processes, particularly at short distances, remains unclear. In addition, for some fusion reactions related to the synthesis of super-heavy nuclei such as $^{64}$Ni+$^{238}$U \cite{Kozu10,Itkis22}, the extracted capture cross sections from the measured mass-total kinetic energy (TKE) distributions at energies above the Bass barrier \cite{Bass74} are significantly smaller than the predicted results of the classic fusion cross-section formula $\sigma_{\rm cap}  = \pi R_B^2(1-V_B/E_{\rm c.m.})$ and the results of ETF+2G approach mentioned above. Considering that the measurements are highly time-consuming for reactions yielding elements 119 and 120, accurate prediction for the capture cross sections with which the evaporation residue (EvR) cross sections \cite{Wang11,Adam20,Nov20,Zhang23,Mad24} can be further estimated, are crucial.

 In addition to the static nuclear potentials, some microscopic dynamics models \cite{MHZ}, such as the time-dependent Hartree-Fock (TDHF) \cite{Maru06,Guo07,Sim14} theory and the improved quantum molecular dynamics (ImQMD) model \cite{ImQMD2014,ImQMD2014a} have also been widely adopted in the study of heavy-ion fusion reactions, with which the time evolution of the densities of the composite system can be self-consistently described. Very recently, the capture thresholds for 144 fusion reactions induced by nearly spherical nuclei have been systematically studied with the TDHF calculations \cite{Yao24}. In conjunction with the  Siwek-Wilczy\'{n}ski (SW) cross-section formula, which incorporates the classic cross-section formula by folding it with a Gaussian barrier distribution, the experimentally measured fusion cross sections at energies around the barriers can be well reproduced for some fusion reactions such as $^{132}$Sn+$^{40,48}$Ca. It is observed that the reaction $Q$-value can influence the fusion cross sections at sub-barrier energies in reactions with nearly spherical nuclei \cite{Yao24}. However, for fusion reactions involving strongly deformed nuclei such as $^{154}$Sm, $^{238}$U and $^{243}$Am, the effective consideration of both the Q-value and nuclear static deformations in the calculation of capture cross sections remains an unresolved issue. Furthermore, considering that the microscopic TDHF calculations are extremely time-consuming, development of a time-saving nucleus-nucleus potential with high accuracy is still necessary for the systematic study of the fusion reactions.

 In this work, we attempt to propose an effective nucleus-nucleus potential based on the Skyrme energy density functional (EDF) and the reaction $Q$-value, for a systematic description of the heavy-ion fusion reactions especially the reaction systems with well deformed nuclei.

 \begin{center}
 	\textbf{ II. EFFECTIVE NUCLEUS-NUCLEUS POTENTIAL }\\
 \end{center}

 In this work, we firstly calculate the frozen nucleus-nucleus potential based on the Skyrme EDF. The entrance-channel nucleus-nucleus potential $V(R)$ between two nuclei is expressed as \cite{liumin,Yao24a}
 \begin{eqnarray}
 	V(R) = E_{\rm tot}(R) - E_1 - E_2,
 \end{eqnarray}
 where $R$ is the center-to-center distance between two fragments. $E_{\rm tot}(R)$ denotes the total energy of the nuclear system, $E_1$ and $E_2$ denote the energies of the reaction partners at an infinite distance, respectively. The total energy of a nuclear system can be expressed as an integral of the Skyrme EDF ${\mathcal H}(\bf{r})$ with the frozen density approximation,
 \begin{eqnarray}
 	E_{\rm tot}(R) =  \int \; {\mathcal H}[ \rho_{1p}({\bf
 		r})+\rho_{2p}({\bf r}-{\bf R}), \rho_{1n}({\bf r})+\rho_{2n}({\bf
 		r}-{\bf R})] \; d{\bf r}. 
 \end{eqnarray}
The energies $E_1$ and $E_2$ are expressed as,
 \begin{eqnarray}
 	E_1 = \int \; {\mathcal H}[ \rho_{1p}({\bf r}), \rho_{1n}({\bf
 		r})] \;
 	d{\bf r}, \\
 	E_2 = \int \; {\mathcal H}[ \rho_{2p}({\bf r}), \rho_{2n}({\bf
 		r})] \; d{\bf r}.
 \end{eqnarray}
 Here, $\rho_{1p}$, $\rho_{2p}$, $\rho_{1n}$ and $\rho_{2n}$ are the frozen proton and neutron densities of the projectile and
 target described by spherical symmetric Fermi functions. In the calculations of energies and corresponding densities of the reaction partners, the Skyrme EDF  with the parameter set SkM*\cite{Bart82} is adopted, utilizing the extended Thomas-Fermi (ETF2) approach to describe both the kinetic energy density and the spin-orbit density within the EDF.

\begin{figure}
	\setlength{\abovecaptionskip}{-0.6cm}
	\includegraphics[angle=0,width=0.8 \textwidth]{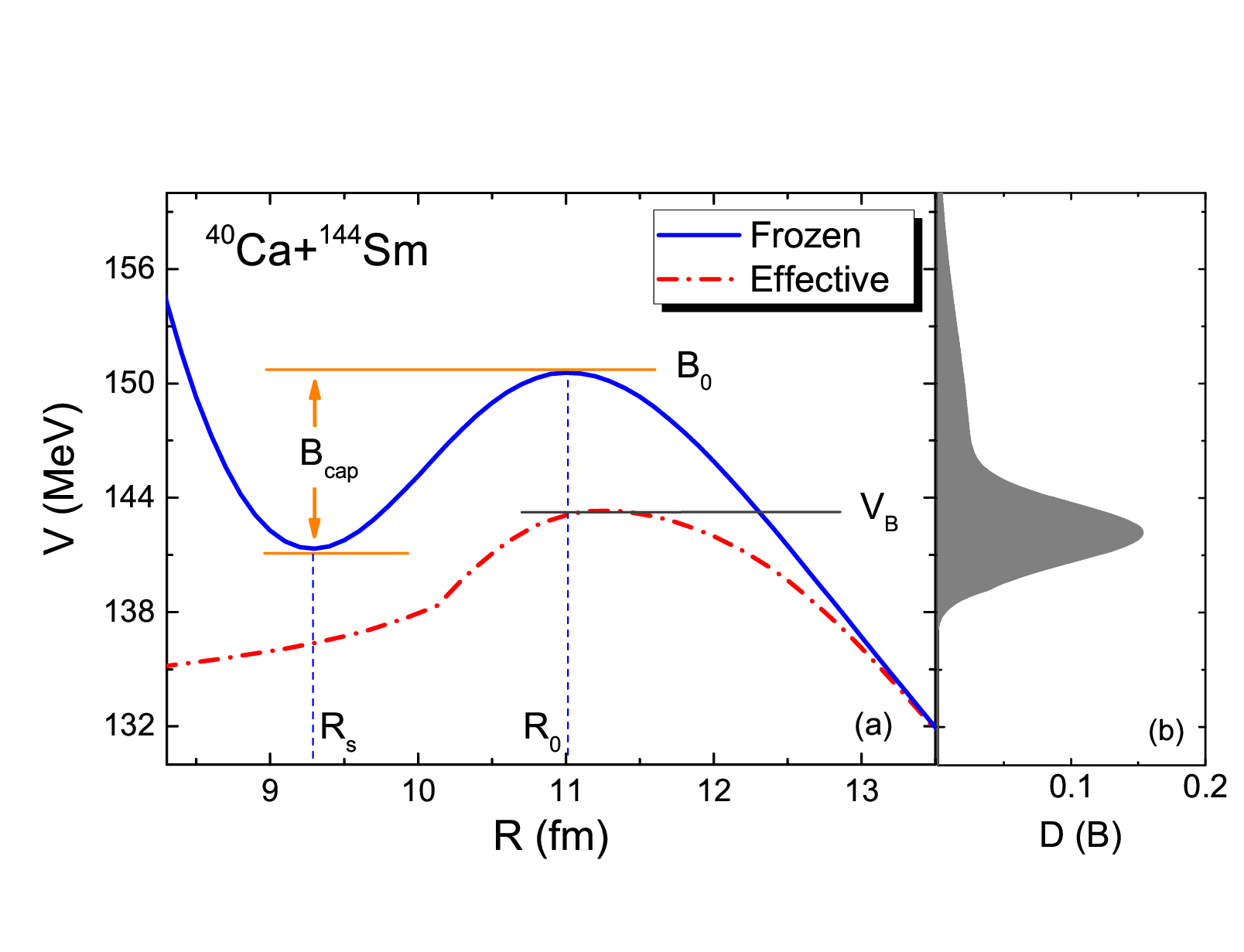}
	\caption{ (a) Nucleus-nucleus potential for $^{40}$Ca + $^{144}$Sm. The solid and the dot-dashed curves denote the frozen and the effective potential, respectively. The short dashed lines denote the position of the frozen barrier $R_0$ and that of the pocket $R_s$. (b) Empirical barrier distribution for $^{40}$Ca + $^{144}$Sm proposed in \cite{liumin}.
 }
\end{figure}

 Based on the entrance nucleus-nucleus potential $V(R)$, the frozen barrier height $B_0$, the depth of the capture pocket $B_{\rm cap}$ can be obtained \cite{Yao24a}. The solid curve in Fig. 1(a) denotes the calculated $V(R)$ for the reaction $^{40}$Ca+$^{144}$Sm. It is known that the frozen potential barrier would be reduced in the fusion process due to the dynamical deformations of the reaction partners and nucleon transfers based on some microscopic dynamics simulations. From the systematic study of  $^{16}$O-induced fusion with the ImQMD model in \cite{ImQMD2014a}, the dynamical barrier height is lower than the frozen one by about $5\%$. 
 
 To consider the influence of the dynamical effect, we propose an effective potential (EP) $V_D (R)$ for fusion system at the regions before two nuclei contact,
 \begin{eqnarray}
 	V_D(R) = V(R)\left[ 1-k \, {\rm erfc}  \left( \frac{R-R_0}{s}-1 \right)  \right] ,
 \end{eqnarray}
 with two parameters $k=0.027$ and $s=1.0$ fm. With the increasing of $R$ between two nuclei, the Coulomb excitation becomes negligible and $V_D(R)$ is consequently close to $V(R)$. The dot-dashed curve in Fig. 1(a) denotes the effective nucleus-nucleus potential for $^{40}$Ca+$^{144}$Sm. The barrier height is reduced from $B_0=150.57$ MeV to $V_B=143.32$ MeV, and at the regions $R>13.5$ fm, one has $V_D(R) \approx  V(R)$. In Fig. 1(b), we show the empirical barrier distribution with a superposition of two Gaussian functions proposed in \cite{liumin}. We note that the value of $V_B$ is close to the average barrier height given by the distribution function and the effective barrier radius $R_B$ is slightly larger than the frozen radius $R_0$.

 After projectile-target contact at energies around the barrier height $V_B$, the frozen density approximation could not be applicable anymore due to the dynamical evolution of the neck. For light fusion system, the compound nucleus would be directly formed after the fusion barrier being overcome since the fission barrier is high enough to  make fission an improbable decay mode at incident energies close to the fusion barrier, and the potential $V_D$ should approach to $-Q$ when the distance between two fragments becomes very small. While for heavy systems such as the reactions leading to SHN, it is thought that the influence of quasi-fission becomes evident. It implies that the effective entrance-channel nucleus-nucleus potential $V_D(R)$ for heavy systems could be significantly larger than the value of $-Q$ at short distances and nucleon transfer through neck becomes an important way to form the compound nucleus as described in the dinuclear system (DNS) model \cite{Adam97,Adam20,Zhang23,Wangnan,Zhu16,Zhang24}. At the regions after projectile-target contact, we write the effective potential (EP) as, 
  \begin{eqnarray}
V_D= V_s+\frac{\Delta U-Q-V_s}{1+\exp[(R-R_2)/s]}+(V_1-V_s)\exp \left( \frac{R-R_1}{s}\right). 
 \end{eqnarray}
 Where $V_s$ and $V_1$ denote the corresponding effective potentials obtained with Eq.(5) at the distance $R=R_B-\Delta R$ and that at the touching point $R=R_1$, respectively. $\Delta R=R_0-R_s$ and $R_2=R_s/2$. $R_0$ and $R_s$ denote the barrier radius and the position of the capture pocket in the frozen nucleus-nucleus potential $V(R)$, respectively (see the dashed lines in Fig. 1).  In this work, the touching point $R_1$ is taken as, 
 \begin{eqnarray}
 	R_1= \left\{
 	\begin{array} {l @{\quad:\quad}l}
 		R_0 &   \Delta R < 1.5 \; {\rm fm}   \\
 		R_0-\Delta R/2 &   \Delta R \geq 1.5  \; {\rm fm}  \\
 \end{array} \right .
\end{eqnarray} 
The quantity $\Delta U=(V_B+Q)-(V_B^{\rm sym}+Q_{\rm sym})$ denotes the difference between the driving potential at the entrance channel and that of the corresponding symmetric system (i.e., the mass asymmetry of the projectile-target combination is about zero).  According to the DNS model, nucleon transfer is mainly governed by the driving potential. Here, we introduce the truncation for fusion reactions, i.e.,  $\Delta U \geq 0$ and $V_s \geq -Q$, considering the energy of the composite system after projectile-target contact should be larger than that of the compound nucleus at its ground state for fusion reactions with heavy nuclei.

 \begin{figure}
 	\setlength{\abovecaptionskip}{ 0.2cm}
 	\includegraphics[angle=0,width=0.85\textwidth]{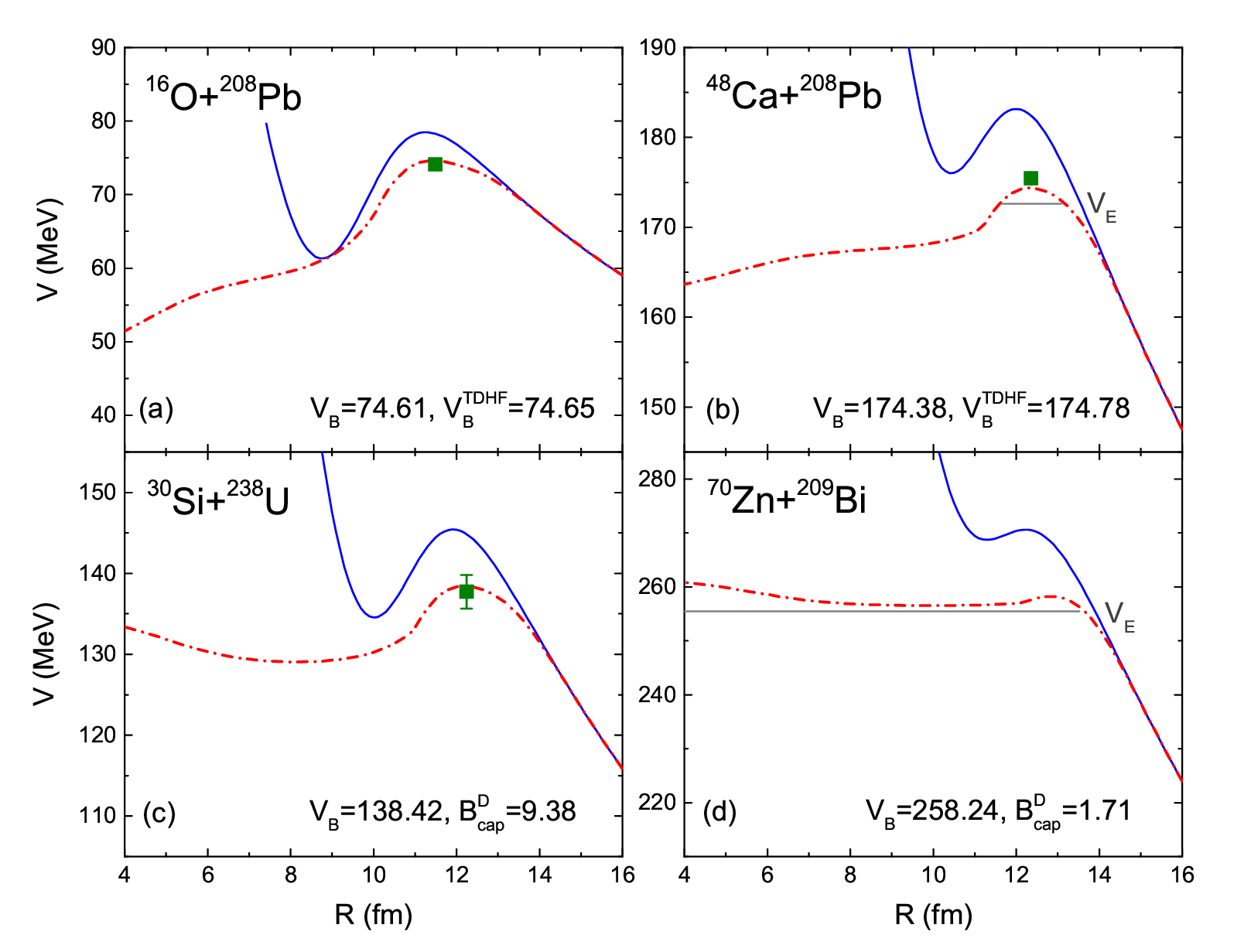}
 	\caption{ The same as Fig. 1(a), but for $^{16}$O + $^{208}$Pb, $^{48}$Ca + $^{208}$Pb, $^{30}$Si+ $^{238}$U and $^{70}$Zn + $^{209}$Bi. The green squares denote the extracted barrier heights \cite{Chen23}. The gray lines in (b) and (d) denote the energies of $V_E=0.99 V_B$. $V_B^{\rm TDHF}$ denotes the predicted capture barrier height in the TDHF calculations \cite{Yao24}.  }
 \end{figure}
 
 In Fig. 2, we show the calculated effective potentials for the reactions  $^{16}$O + $^{208}$Pb,  $^{48}$Ca + $^{208}$Pb, $^{30}$Si + $^{238}$U and $^{70}$Zn + $^{209}$Bi. The frozen nucleus-nucleus potentials (solid curves) are also presented for comparison. In Ref. \cite{Chen23}, the fusion barrier parameters for 367 reaction systems are systematically extracted, based on 443 datasets of measured fusion/fission cross sections. The green squares denote the extracted barrier heights \cite{Chen23}. One sees that the capture barrier heights $V_B$ from the effective potential (EP) are in good agreement with the experimental ones. In this work, the calculated $V_B$ with the proposed EP are also systematically compared with the extracted barrier heights. We find that the root-mean-square (rms) deviations with respect to the 443 extracted barrier heights is 1.53 MeV, which is smaller than the Bass \cite{Bass74,Bass80} and BW91 \cite{BW91} potentials. For $^{16}$O + $^{208}$Pb and $^{48}$Ca + $^{208}$Pb, we also note that the results of EP are very close to the corresponding TDHF capture thresholds \cite{Yao24}. The rms deviations with respect to the 144 capture barrier heights \cite{Yao24} predicted with the TDHF calculations is only 1.05 MeV. Considering that the microscopic TDHF calculations are extremely time-consuming, the proposed effective nucleus-nucleus potentials with similar accuracy would be quite useful for the systematic study of the fusion reactions.

 From Fig. 2, one sees that for the intermediate fusion system $^{16}$O + $^{208}$Pb, the EP evidently decreases with the decreasing of  $R$ after projectile-target contact and gradually approaches to the value of $-Q=46.5$ MeV. Whereas for the heavy fusion system $^{70}$Zn + $^{209}$Bi, the EP just slightly decreases by $B_{\rm cap}^D=1.71$ MeV after projectile-target contact and at very short distances the potentials are even higher than $V_B$ by about 2 MeV, which implies that the formation of the compound nuclei is a relatively slow process and the competition among fusion, quasi-fission and deep-inelastic scattering would be evident for this system. The gray lines in (b) and (d) denote the energies $V_E=0.99 V_B$ below which the elastic scattering is a dominant process. From Fig. 2(d), we also note that it could be quite difficult to form the compound nuclei in the reaction  $^{70}$Zn + $^{209}$Bi at energies lower than $V_E$ based on the barrier penetration concept, since the potential barrier becomes extremely thick. In addition, comparing with the EP for $^{30}$Si + $^{238}$U in which the depth of the capture pocket $B_{\rm cap}^D=9.38$ MeV is much larger than that of $^{70}$Zn + $^{209}$Bi, the quasi-fission and deep-inelastic scattering events would be much more in $^{70}$Zn + $^{209}$Bi at energies around the capture barriers.
 
 \newpage
     
  \begin{center}
 	\textbf{ III. CAPTURE CROSS SECTIONS }\\
 \end{center}
 
 It is known that the couplings between the relative motion of colliding nuclei and the intrinsic degrees of freedom play an important role for heavy-ion fusion reactions. To consider these couplings, Stelson introduced a distribution of barrier heights $D(B)$ in the calculation of the fusion excitation function \cite{Stel88}. A well-known example is the Gaussian distribution of barrier heights predicted from different orientations of colliding nuclei undergoing slow deviations from sphericity \cite{Stel88}. In Ref. \cite{SW04}, an analytical cross section formula is proposed  for describing the capture excitation function at energies around the Coulomb barrier by Siwek-Wilczy\'{n}ska and Wilczy\'{n}ski (SW) under the Gaussion distribution assumption,
 \begin{equation}
 	\sigma_{\rm {cap}}(E_{\rm c.m.})=\pi R_m^{2}  \frac{W}{\sqrt{2}E_{\rm c.m.}}
 	[X  {\rm erfc}(-X)+\frac{1}{\sqrt{\pi}}\exp(-X^{2}) ],
 \end{equation}
 where $X = (E_{\rm c.m.}-V_B)/\sqrt{2}W $. $V_B$ and $W$ denote the centroid and the standard deviation of the Gaussian function, respectively. $R_m$ denotes the average barrier radius which is usually set as $R_m=R_B$. In this work, the average barrier radius,  
  \begin{equation}
 	R_m= \frac{\int  (V_D-V_E)   R \, dR}{\int  (V_D-V_E)  \, dR}     
 \end{equation}
is calculated with the effective potential $V_D$ at energies around the capture barrier height $E_{\rm c.m.}= (1 \pm 0.01) V_B $. We note that  $R_m\approx R_B$ for most of fusion systems (see the positions of the green squares in Fig. 2). Whereas for the reaction systems with very shallow capture pockets, the value of $R_m$ would be significantly smaller than $R_B$, which will be discussed later.

 For fusion reactions induced by heavy target nuclei with a quadrupole deformation of $\beta_2$, the range of barrier heights $\Delta B \propto \beta_2 V_B  R_T/R_B$  due to different orientations of the deformed nuclei can be derived with average target radius $R_T$ \cite{Das98}. The $V_B$ and $\beta_2$ dependence of $W$ is therefore expected for reactions with deformed nuclei. On the other hand, the excitation energy of the compound nuclei (related to the reaction $Q$-value)  at energies around the barrier in reaction with deformed nuclei is usually larger than that with neighboring spherical nuclei. For example, the $Q$-value of the reaction $^{16}$O+$^{154}$Sm is higher than that of $^{16}$O+$^{144}$Sm by about 12 MeV (which will be further discussed later). In addition, for heavy nucleus with large deformations, the excitation threshold $\varepsilon_{th}$, which is defined as the energy of the lowest excited state of the reaction partners, is usually very small. For example, the value of $\varepsilon_{th} =0.082$ MeV for $^{154}$Sm and $\varepsilon_{th} =0.045$ MeV for $^{238}$U. 
 Considering that the deformation parameter $\beta_2$ is model dependent \cite{Wang19,Wangbing}. It could be interesting to modify the barrier height effects with $W$ having a $Q$-value and excitation threshold dependence.

Recently, the SW formula has been applied to study the fusion excitation functions for reactions involving nearly spherical nuclei, with the three parameters determined by the TDHF calculations \cite{Yao24}. It has been found that the standard deviation of the Gaussian function $W$ is related to the reaction $Q$-value and deformation effects. A relatively higher excitation energy at the capture position can lead to stronger impacts for dynamical deformations and nucleon transfer during the capture process, thereby broadening the width of the barrier distribution. Following the idea in Ref. \cite{Yao24}, we extend the expression of $W$ to intermediate and heavy fusion reactions induced by nuclei with large deformations. The value of $W$ for the fusion reactions induced by nuclei with large deformations is parameterized as,
\begin{equation}
	W=\frac{2}{3}W_0 +\frac{1}{3}W_1
\end{equation} 
with $W_0=0.052(V_B+Q)$ for the reactions induced by nearly spherical nuclei \cite{Yao24}. From a systematic study of the fusion barrier parameters, Chen et al. find that the value of $W \approx (0.014+0.135\lambda{}_B ) V_B$ with the reduced de Broglie wavelength $\lambda_B=\hbar/\sqrt{2\mu V_B}$ \cite{Chen23}. Considering the systematic behavior of $W$, we write $W_1=(0.048+\eta \, \varepsilon_{\rm 1}/\varepsilon_{\rm 2}) V_B$. Here, $\eta=|A_1-A_2|/(A_1+A_2)$ denotes the mass asymmetry of the reaction system with $A_1$ and $A_2$ being the mass number of the projectile and that of the target, respectively. $\varepsilon_{\rm 1}$ and $\varepsilon_{\rm 2}$ denote the smaller value of the energies of the lowest excited states for the reaction partners and that of the larger one, respectively. Simultaneously, we introduce a truncation for the value of $W$, i.e.,  
$W \leq 2.5 W_0$. From Eq.(10), one can see that the $V_B$ dependence of barrier distribution width is considered in $W_1$ and the deformation effects are indirectly considered with the $Q$-value in $W_0$ and the energies of the lowest excited states in $W_1$.

In Table I, we list the calculated barrier parameters for some reactions under consideration. The barrier height $V_B$ and average barrier radius $R_m$ are obtained with Eq.(5) and Eq.(9), respectively. The standard deviation of the Gaussian function $W$ is obtained with Eq.(10). The reaction $Q$-value for unmeasured super-heavy system is from the prediction of the Weizs\"acker-Skyrme (WS4) mass model \cite{WS4} with which the known masses can be reproduced with an rms error of $\sim0.3$ MeV \cite{Zhao22} and the known $\alpha$-decay energies of SHN can be reproduced with deviations smaller than 0.5 MeV \cite{Ogan15,Wang15}.

 \begin{table}
 	\caption{ Barrier parameters with the effective potential. $Q$ denotes the reaction $Q$-value.}
 	\begin{tabular}{ccccc}
 		\hline\hline
 		
 		~~~reaction~~~  & ~~~$V_B$ (MeV)~~~ & ~~~W (MeV)~~~ & ~~~$R_m$ (fm)~~~ & ~~~$Q$ (MeV)~~~   \\
 		\hline
 		$^{16}$O+$^{144}$Sm  &  60.51   &  1.66   &  10.66  &  $-28.54$     \\
 		$^{16}$O+$^{154}$Sm  &  59.42   &  2.66   &  10.87  &  $-16.43$   	\\
 		$^{32}$S+$^{154}$Sm  &  114.34  &  4.61   &  11.31  &  $-60.60$      \\
 		$^{48}$Ca+$^{154}$Sm &  137.93  &  4.36   &  11.75  &  $-90.74$      \\
 		$^{30}$Si+$^{238}$U  &  138.42  &  4.51   &  12.24  &  $-93.01$       \\
 		$^{48}$Ca+$^{238}$U  &  191.19  &  3.95   &  12.71  &  $-160.78$       \\
 		$^{52}$Cr+$^{232}$Th &  225.26  &  4.92   &  8.15   &  $-187.40$      \\
 		$^{52}$Cr+$^{248}$Cm &  237.90  &  4.41   &  6.13   &  $-203.95$      \\
 		$^{54}$Cr+$^{243}$Am &  235.49  &  3.68   &  5.98   &  $-207.20$      \\ 		
 		$^{64}$Ni+$^{238}$U  &  263.37  &  3.22   &  5.42   &  $-238.61$      \\		
 		\hline\hline
 	\end{tabular}
 \end{table}

\begin{figure}
	\setlength{\abovecaptionskip}{ -2cm}
	\includegraphics[angle=0,width=0.95\textwidth]{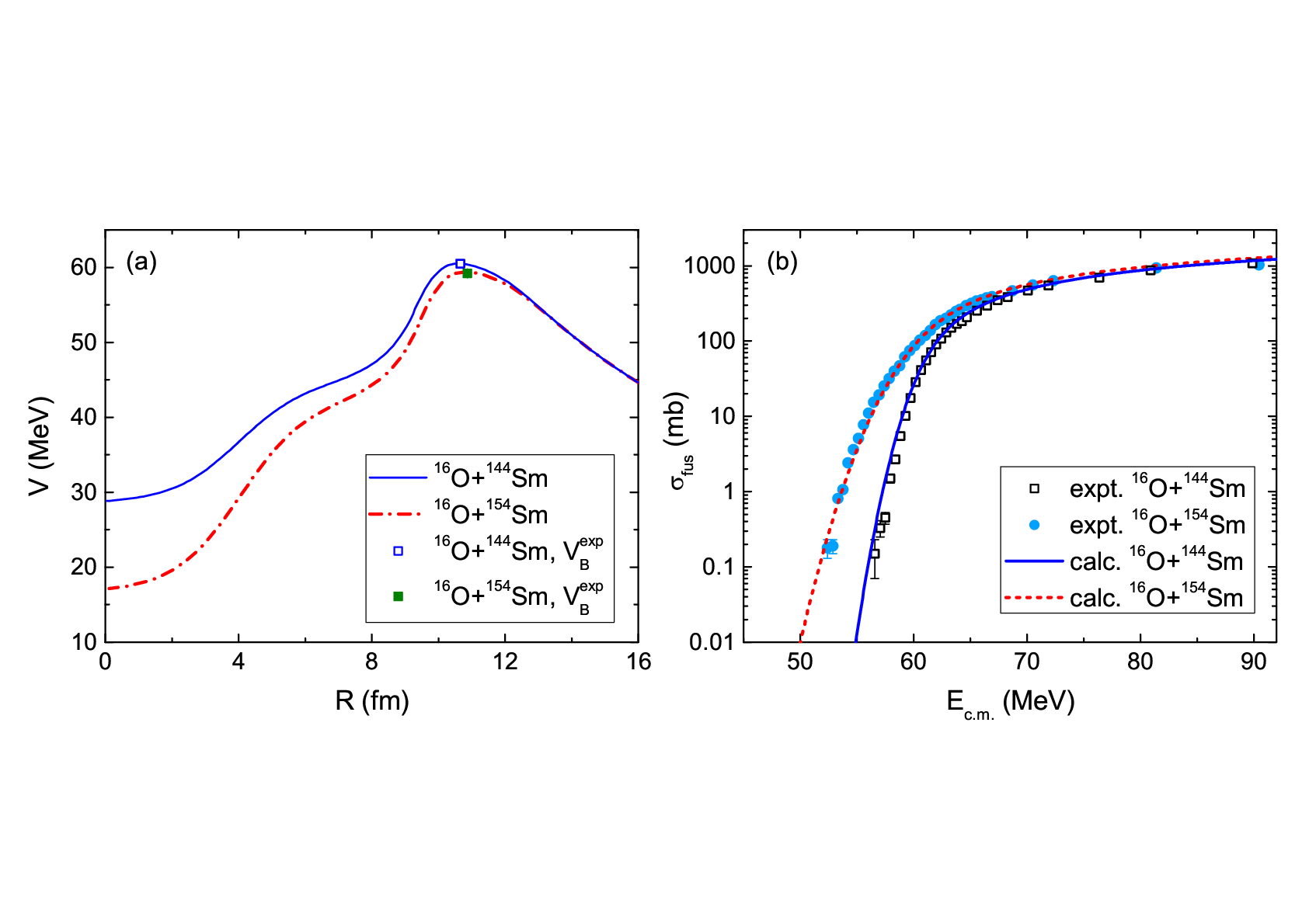}
	\caption{ (a) The same as Fig. 2, but for  $^{16}$O + $^{144,154}$Sm. The data for the barrier heights are taken from \cite{Chen23}.  (b) Fusion excitation functions for  $^{16}$O + $^{144,154}$Sm. The scattered symbols denote the experimental data taken from \cite{Lei95} and the curves denote the predicted results with Eq.(8).  }
\end{figure}

\begin{figure}
	\setlength{\abovecaptionskip}{ -0.2 cm}
	\includegraphics[angle=0,width=0.9\textwidth]{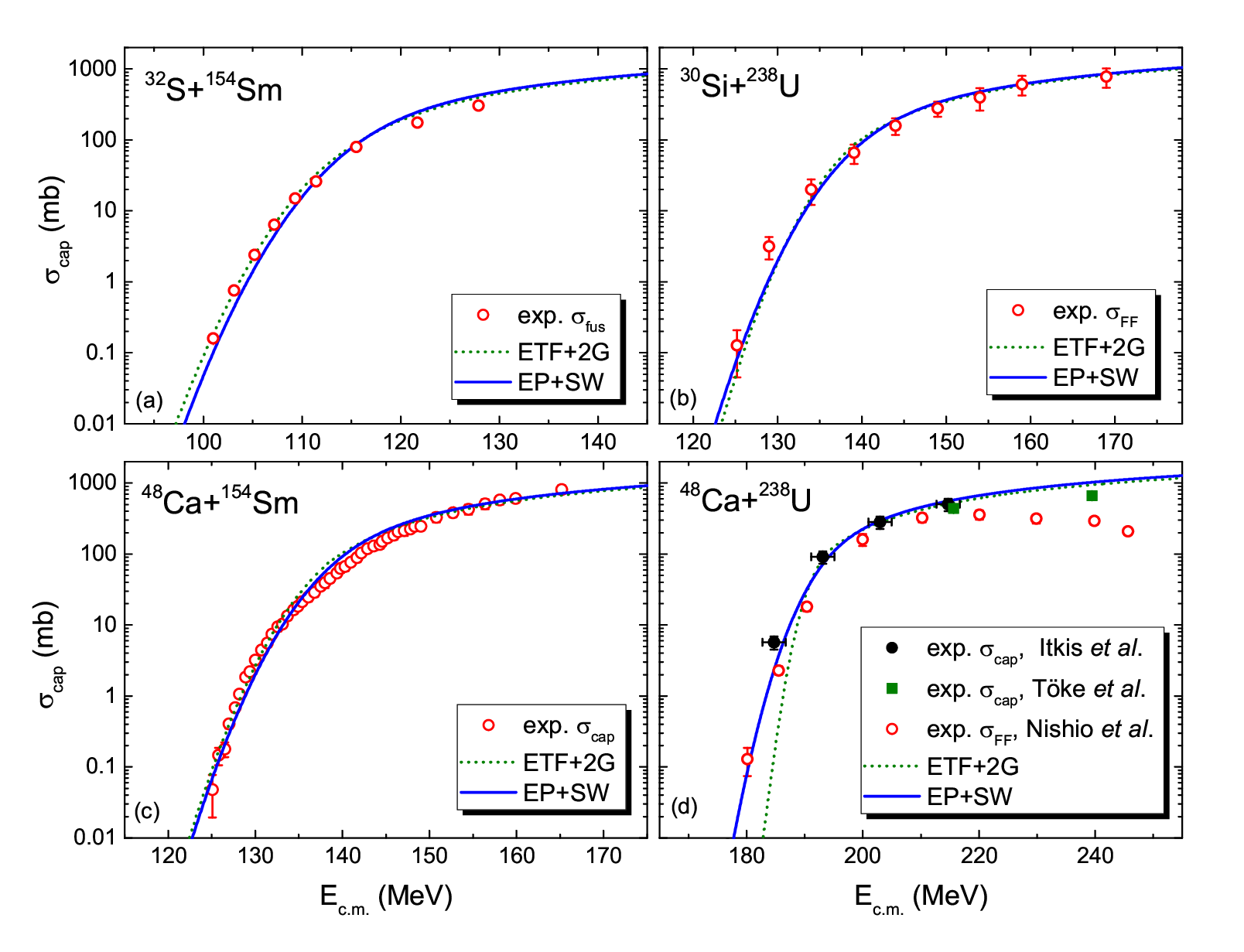}
	\caption{ Capture excitation functions for the reactions $^{32}$S + $^{154}$Sm \cite{Gom94},  $^{30}$Si + $^{238}$U \cite{Nishio10},  $^{48}$Ca + $^{154}$Sm \cite{Trot05}, and  $^{48}$Ca + $^{238}$U \cite{Nishio12,Itkis22,Toke85}. The solid curves and the short dashes denote the results of the effective potential together with the SW formula given by Eq.(8) and the results of the empirical two Gaussian (2G) barrier distribution approach \cite{liumin}, respectively. }
\end{figure}
     
In Fig. 3, we show the effective potentials and the fusion excitation functions for the reactions $^{16}$O + $^{144,154}$Sm. One sees that in Fig. 3(a) the calculated barrier heights are in good agreement with the experimental ones. The fusion barrier height for the neutron-rich system $^{16}$O + $^{154}$Sm is lower than that of $^{16}$O + $^{144}$Sm by 1.09 MeV. At very short distances, the potentials approach to the corresponding values of $-Q$. From Fig. 3(b), One sees that the measured fusion cross sections for both $^{16}$O + $^{144}$Sm and $^{16}$O + $^{154}$Sm can be well reproduced.

 In Fig. 4, we show the calculated capture excitation functions for the reactions with deformed nuclei $^{154}$Sm and $^{238}$U. The scattered symbols denote the experimental data. The solid curves and the short dashes denote the results of the effective potential together with the SW formula given by Eq.(8) and the results of  ETF+2G  approach \cite{liumin}, respectively. We note that at energies around the capture barriers the experimental data can be reasonably well reproduced by both of the two approaches. For $^{48}$Ca + $^{238}$U, the measured cross sections at sub-barrier energies are reproduced much better with the EP+SW approach. Here, we would like to emphasize that the SW formula is obtained from the folding of the Gaussian barrier distribution and the classic over-barrier fusion cross-section expression which is not applicable for deep sub-barrier energies since it does not take into account the barrier penetration effect. From Fig. 3 and Fig. 4, one can see that the measured capture cross-sections at energies around the barrier can be well reproduced by the SW formula, which indicates the folding of a suitable barrier distribution plays a role for describing the capture cross sections at around barrier energies. In addition, the Wong formula \cite{Wong73} which considers the quantum-mechanical tunneling effect is used by folding the two-Gaussian barrier distribution in the ETF+2G calculations. One notes that the results from the two approaches are close to each other for the reactions  $^{32}$S + $^{154}$Sm,  $^{30}$Si + $^{238}$U and $^{48}$Ca + $^{154}$Sm, which also indicates the importance of the barrier distribution.

    \begin{figure}
        \setlength{\abovecaptionskip}{-0.2cm}
        \includegraphics[angle=0,width=0.7\textwidth]{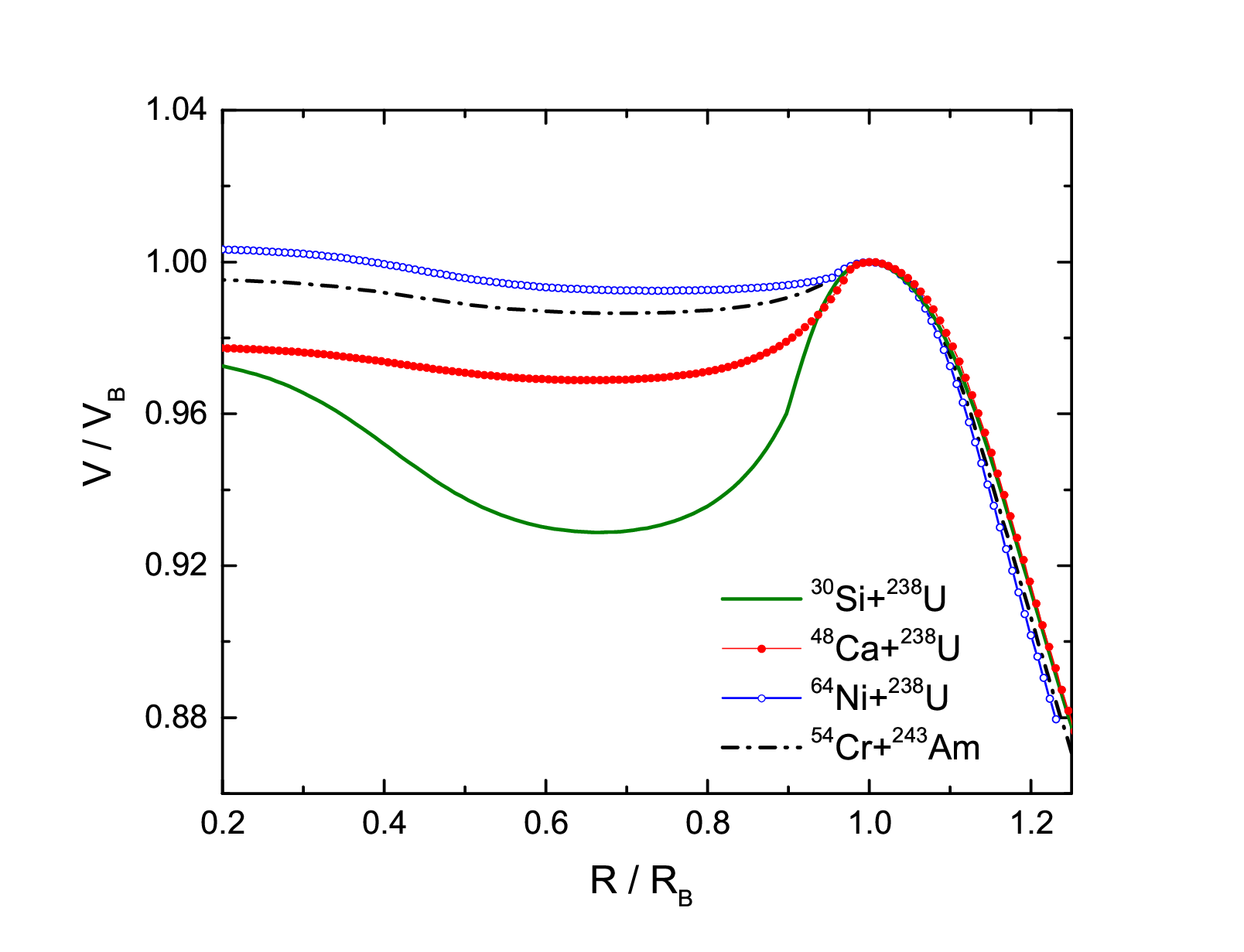}
        \caption{ Effective nucleus-nucleus potentials for $^{30}$Si,$^{48}$Ca,$^{64}$Ni + $^{238}$U and  $^{54}$Cr+$^{243}$Am. }
    \end{figure}

\begin{figure}
	\setlength{\abovecaptionskip}{ 0.2cm}
	\includegraphics[angle=0,width=0.95\textwidth]{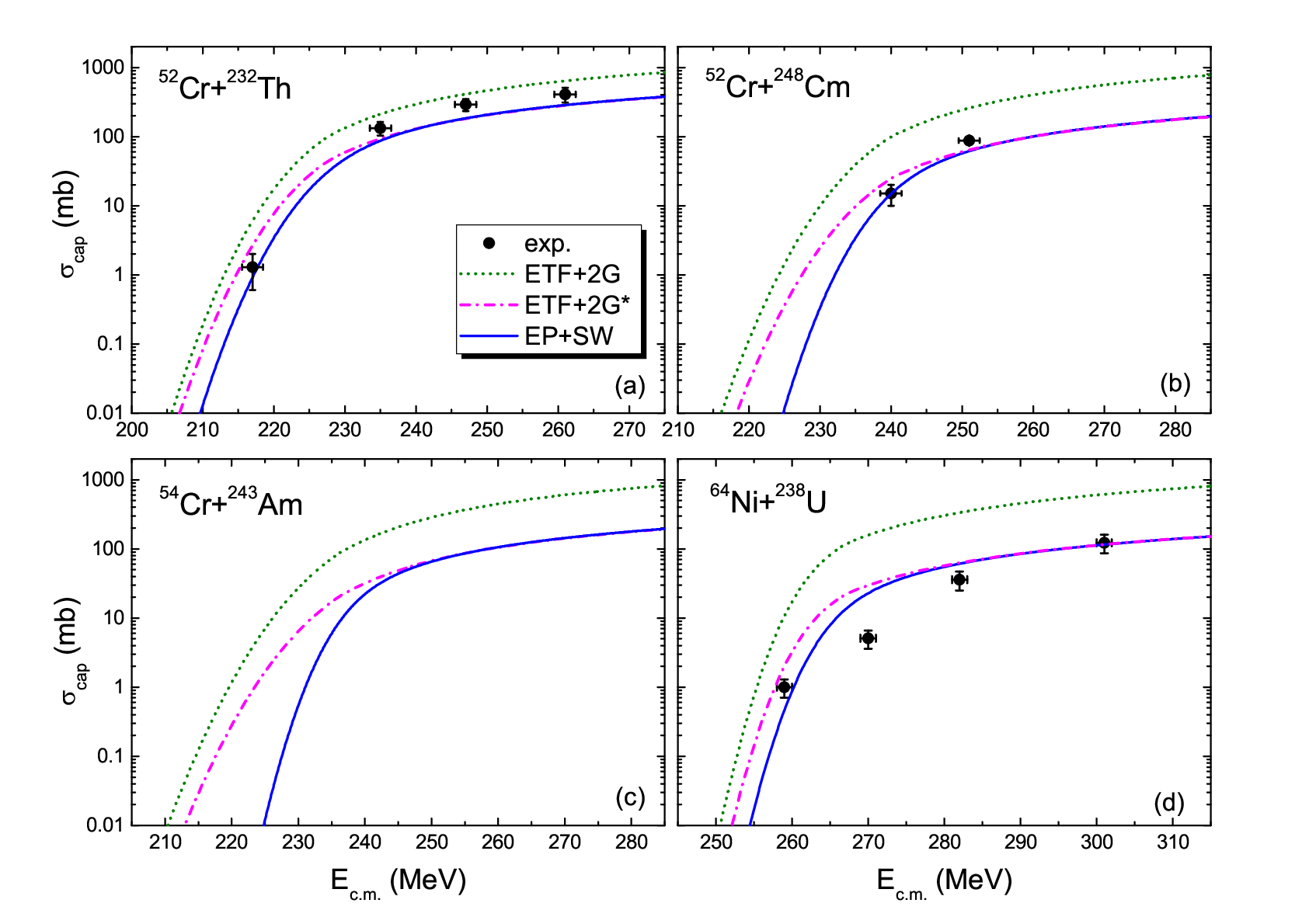}
	\caption{(Color online) The same as Fig. 4, but for $^{52}$Cr+$^{232}$Th,$^{248}$Cm, $^{54}$Cr+$^{243}$Am and $^{64}$Ni+$^{238}$U. The dot-dashed curves denote the results with the 2G approach but with the average barrier radius $R_m$ rather than $R_B$ in the calculations. The data are taken from \cite{Itkis22}.}
\end{figure}

Simultaneously, the proposed effective potential (EP) is also tested for describing some fusion reactions leading to the synthesis of super-heavy nuclei. In Fig. 5, we show the calculated EP for the reactions $^{30}$Si,$^{48}$Ca,$^{64}$Ni + $^{238}$U and  $^{54}$Cr+$^{243}$Am. To see the depth of the capture pockets more clearly, the potential and the distance are scaled by $V_B$ and $R_B$, respectively. We note that the depths of the capture pockets significantly decrease with the increasing of the product of projectile-target charge numbers $Z_1 Z_2$. Very recently, Yao et al. find that the yields of fission-like events in the measured Mass-TKE distributions  and the ratio of capture to deep-inelastic scattering events evidently decreases with the decreasing of the depths of the capture pockets $B_{\rm cap}$ \cite{Yao24a}. 
To understand the physics behind, we study the average barrier radius $R_m$. From Table I, one notes that for the reactions induced by $^{52,54}$Cr and $^{64}$Ni the average barrier radii are significantly smaller than those of other reactions although the contact distances are larger. With smaller average barrier radii for the reactions with $^{52,54}$Cr and $^{64}$Ni, the corresponding capture cross sections at above barrier energies are expected to be reduced. In Fig. 6, we show the predicted capture cross sections for the reactions $^{52}$Cr+ $^{232}$Th, $^{52}$Cr+ $^{248}$Cm,  $^{54}$Cr+$^{243}$Am and $^{64}$Ni+ $^{238}$U. The green dashed and the blue solid curves denote the predicted results with the empirical 2G barrier distribution approach and those with the EP+SW approach, respectively. The solid circles denote the experimental data taken from \cite{Itkis22}. One sees that for $^{52}$Cr+ $^{248}$Cm and $^{64}$Ni+ $^{238}$U, the capture cross sections are significantly over-predicted by the ETF+2G approach. The dot-dashed curves denote the results with the ETF+2G approach but adopting the average barrier radius $R_m$ in the calculations. With $R_m$ rather than $R_B$ in the calculations, the capture cross sections at above barrier energies are much better reproduced. The predicted capture cross sections at sub-barrier energies with the effective potential are evidently smaller than those with the 2G approach for $^{52,54}$Cr induced reactions, which implies that the predicted 2n EvR cross sections would be much smaller than those with ETF+2G approach for the reaction $^{54}$Cr+$^{243}$Am.

\begin{center}
	\textbf{IV. SUMMARY}
\end{center}

Based on the frozen nucleus-nucleus potential from the Skyrme energy density functional together with the extended Thomas-Fermi approach and sudden approximation for densities, we propose an effective approach to obtain the capture barrier heights and average barrier radii for heavy-ion fusion reactions. The 443 extracted  barrier heights can be well reproduced with an rms error of 1.53 MeV. We also note that the results of effective potentials are very close to the corresponding TDHF capture thresholds for $^{16}$O + $^{208}$Pb and $^{48}$Ca + $^{208}$Pb. The rms deviations with respect to the 144 capture barrier heights predicted with the TDHF calculations is only 1.05 MeV, which provides us with a useful balance between accuracy and computation cost allowing a large number of fusion systems with a simple uniform manner. Together with Siwek-Wilczy\'{n}ski formula in which the three parameters are determined by the proposed effective potentials, the measured capture cross sections at energies around the barriers can be well reproduced for a series of fusion reactions induced by both spherical and well deformed nuclei. The shallow capture pockets and small values of the average barrier radii play a role in the reduction of the capture cross sections for $^{52,54}$Cr and $^{64}$Ni induced reactions which are related to the synthesis of super-heavy nuclei. It seems that the decreasing of the capture pocket depth for heavy reaction system results in the increase of the deep-inelastic collisions at energies around the capture barrier, which consequently suppresses the production of compound system.

\newpage

\begin{center}
	\textbf{ACKNOWLEDGEMENTS}
\end{center}
This work was supported by National Natural Science Foundation of
China (Nos. 12265006, 12375129, U1867212), Guangxi Natural Science Foundation (2017GXNSFGA198001). Some data tables about the capture barrier heights are available at http://www.imqmd.com/fusion/

\end{document}